\documentclass[%
 reprint,
superscriptaddress,
 amsmath,amssymb,
 aps,
showkeys
]{revtex4-2}

\usepackage{tabularx}
\usepackage{graphicx}% Include figure files
\usepackage{dcolumn}% Align table columns on decimal point
\usepackage{bm}% bold math
\usepackage{braket}
\usepackage[colorlinks, allcolors=blue]{hyperref}
\begin{document}

%\preprint{APS/123-QED}

\title{Anisotropic Superconductivity with Enhanced Critical Field in Strained RuO$_{2}$ }

\author{Younes Ghorbani}
\altaffiliation{These authors contributed equally to this work}
\affiliation{Department of Materials Science and Engineering, The Ohio State University, Columbus, Ohio 43210 USA}
\author{Samuel J. Poage}
\altaffiliation{These authors contributed equally to this work}
\affiliation{Department of Materials Science and Engineering, The Ohio State University, Columbus, Ohio 43210 USA}
\author{Xueshi Gao}
\altaffiliation{These authors contributed equally to this work}
\affiliation{Department of Physics, The Ohio State University, Columbus, Ohio 43210 USA}
\author{Luka Mitrovic}
\affiliation{Department of Materials Science and Engineering, Cornell University, Ithaca, NY 14853, USA}
\author{Neha Wadehra}
\affiliation{Department of Materials Science and Engineering, Cornell University, Ithaca, NY 14853, USA}
\author{Benjamin Z. Gregory}
\affiliation{Department of Materials Science and Engineering, Cornell University, Ithaca, NY 14853, USA}
\author{Suchismita Sarker}
\affiliation{Cornell High Energy Synchrotron Source, Cornell University; Ithaca, NY 14853, USA}
\author{Bet\"{u}l Pamuk}
\affiliation{Department of Physics and Astronomy, Williams College, Williamstown, Massachusetts 01267, USA}
\author{Andrej Singer}
\affiliation{Department of Materials Science and Engineering, Cornell University, Ithaca, NY 14853, USA}
\author{Darrell G. Schlom}
\affiliation{Department of Materials Science and Engineering, Cornell University, Ithaca, NY 14853, USA}
\affiliation{Kavli Institute at Cornell for Nanoscale Science, Cornell University, Ithaca, New York 14853, USA}
\affiliation{Leibniz-Institut für Kristallzüchtung, Max-Born-Str. 2, 12489 Berlin, Germany}
\author{Chun Ning Lau}
\affiliation{Department of Physics, The Ohio State University, Columbus, Ohio 43210 USA}
\author{Kaveh Ahadi}
 \email{ahadi.4@osu.edu}
\affiliation{Department of Materials Science and Engineering, The Ohio State University, Columbus, Ohio 43210 USA}
\affiliation{Department of Electrical and Computer engineering, The Ohio State University, Columbus, Ohio 43210 USA}

\date{\today}% It is always \today, today,

\begin{abstract}
\section*{Abstract}
The superconductivity in RuO$_{2}$ emerges under strain. RuO$_{2}$ is also an altermagnet candidate. The nature of superconductivity and its relation with neighboring orders, however, are not understood. To address this problem, we grew epitaxial RuO$_{2}$ films on TiO$_{2}$(100) and TiO$_{2}$(110) single crystal substrates and studied the electronic transport and emergent superconductivity along various crystallographic directions. We show that the superconducting transition strongly depends on the growth orientation and the crystallographic direction of the transport in the RuO$_{2}$ films. We also observe a strong violation of Pauli paramagnetic limit with in-plane applied magnetic field ($H_{c2,\parallel}/H_p\sim5.5$) which we attribute to strong spin-orbit scattering. These results offer opportunities for epitaxially engineered superconductors.
\end{abstract}

\keywords{superconductivity, epitaxial strain, RuO$_{2}$}%Use show keys class option if keyword
\maketitle

%\tableofcontents

\section{Introduction}

The ability to precisely tune superconductivity is highly desirable for fundamental understanding and any application of superconductors. For example, pressure control of superconductivity has sparked interest due to its extraordinary ability to manipulate the superconducting order parameter \cite{shimizu2002superconductivity, drozdov2015conventional, PhysRevLett.78.118, RevModPhys.90.011001}. 

Epitaxial strain in superconducting films and heterostructures provides a tuning knob that is compatible with a wide range of probes and device architectures. Epitaxial strain has exhibited success in enhancing superconductivity in various known superconductors, including cuprates \cite{locquet1998doubling, bozovic2002epitaxial}, titatanes \cite{ahadi2019enhancing, russell2019ferroelectric}, and iron-based superconductors \cite{engelmann2013strain}. Epitaxial strain control of superconductivity, however, has remained limited to known superconductors.

RuO$_{2}$ has a rutile crystal structure (P4$_{2}$/mnm) in which each ruthenium has four electrons occupying its 4$d$ $t_{2g}$ manifold. Ruthenium atoms are octahedrally coordinated with six oxygen atoms, where the equatorial bonds are longer than the apical bonds. This breaks the degeneracy of the xy, xz, and yz orbitals in the $t_{2g}$ manifold, pushing the xy orbital down and the xz/yz orbitals up in energy. Accordingly, the xy orbital is full, and the xz/yz orbitals remain half-filled, which results in a bump in the density of states below the Fermi level (Figure 1a) \cite{xu1989self, ruf2021strain}.

The emerging superconductivity in RuO$_{2}$ is highly sensitive to strain. While bulk RuO$_{2}$ does not exhibit superconductivity, films strained to TiO$_{2}$(100) and (110) substrates exhibit superconductivity with critical temperature as high as 1.7 K \cite{ruf2021strain, wadehra2025strain, uchida2020superconductivity}. RuO$_{2}$ films grown on MgF$_{2}$\cite{uchida2020superconductivity}, yttria-stabilized zirconia (111)\cite{fatima2021systematic}, $\alpha$-Al$_{2}$O$_{3}$(0001)\cite{fatima2021systematic} and $\alpha$-Al$_{2}$O$_{3}$(1$\bar1$02)\cite{nunn2021solid} single-crystal substrates do not show a superconducting transition. It is not clearly understood why epitaxial strain gives rise to superconductivity but, in the RuO$_{2}$ films strained to TiO$_{2}$(110), the equatorial-to-apical bond length ratio is quenched, enhancing the Fermi level density of states (Figure 1b and c). Furthermore, the epitaxial strain modifies the lattice dynamics, softening two phonon modes along [001] direction, which is predicted to play a role in mediating ruthenium 4$d$ electron pairing \cite{uchida2020superconductivity}. Finally, proximity to an itinerant antiferromagnetic ordering \cite{jeong2025metallicity} could play a role in emergent superconductivity in strained RuO$_{2}$.

\begin{figure*}[t!]
\includegraphics[width=2\columnwidth]{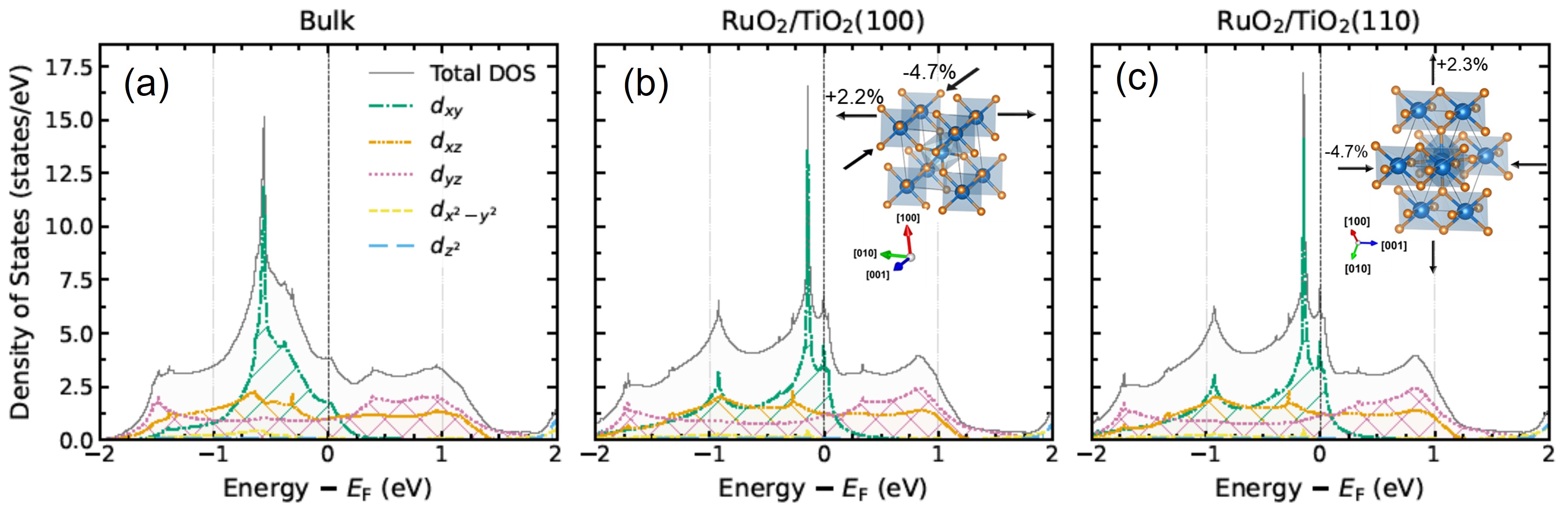}
 \caption{\textbf{The density of states (DOS) of bulk (unstrained) and epitaxially strained RuO$_2$.} DOS calculations are from first principles for (a) unstrained RuO$_2$ and commensurately strained RuO$_2$ to (b) TiO$_2(100)$ and (c) TiO$_2(110)$ single crystal substrates. Inset exhibits the epitaxial strain geometry schematically. Focusing on the DOS around the Fermi level, we observe that the main contribution to the electronic structure originates from the Ru $t_{2g}$ states (i.e $d_{xz}$, $d_{yz}$, and $d_{xy}$ $d$-orbitals), shown with dotted lines. The Ru $e_g$ states, shown with dashed lines, lie about 2 eV above the Fermi level. We focus here on the narrow peak in the DOS that lies about 0.5 eV below the Fermi level. This peak is dominated by the Ru 4$d_{xy}$ states. While the exact position of this peak depends on the exchange and correlation (XC) functional of choice, this peak qualitatively shifts closer to the Fermi level in both strained RuO$_2(100)$ and strained RuO$_2(110)$ films regardless of the choice of approximation \cite{ruf2021strain, wadehra2025strain}. Upon these strain configurations, the energy shift of the Ru 4$d_{xy}$ states towards the Fermi level causes an enhancement of the Fermi DOS, making RuO$_2$ an ideal candidate for strain engineering of electronic instabilities.}
    \label{fig:Fig1}
\end{figure*} 

Neutron diffraction resolved an itinerant antiferromagnetic ordering in bulk RuO$_{2}$ \cite{berlijn2017itinerant}. More recently, RuO$_{2}$ was proposed as an altermagnetic candidate with enhanced spin splitting \cite{vsmejkal2022emerging}. The absence of phase transition signature in the electrical resistivity \cite{lin2004low, glassford1994electron}, magnetic susceptibility \cite{guthrie1931magnetic, fletcher1968magnetic, ryden1970magnetic}, and heat capacity \cite{cordfunke1989thermophysical, o1997gibbs} experiments, however, suggests a Pauli paramagnetic ground state in the stoichiometric bulk RuO$_{2}$. Theoretical calculations suggest that a magnetic ground state would be unfavorable in stoichiometric RuO$_{2}$ due to the low density of states and dispersive bands at the Fermi level \cite{smolyanyuk2024fragility}. The same study predicts that hole doping, e.g., ruthenium vacancies, could shift the Fermi level towards the xy orbital bump in the density of states (see Figure 1a), stabilizing an itinerant antiferromagnetic ordering within a reasonable Hubbard $U$ range. Similarly, the predicted strain control of the Fermi-level density of states in RuO$_{2}$ (Figure 1b and c) could engineer an itinerant magnetic ordering \cite{stoner1938collective, jeong2025metallicity}.

While the ground state of the stoichiometric bulk RuO$_{2}$ is a trivial Pauli paramagnet, epitaxial strain could tune the lattice dynamics, modify the electronic structure, and give rise to electronic instabilities, including superconductivity. Here, we grew epitaxial RuO$_{2}$ films on TiO$_{2}$(100) and TiO$_{2}$(110) single crystals and studied electronic transport. We observe a robust superconductivity, violating Pauli limit with $H_{c2,\parallel}/H_p$ ratio reaching $\sim5.5$. We also observe a strong dependence of superconducting parameters on the substrate orientation, epitaxial strain, and crystallographic direction of electronic transport.

\section{Results and Discussion}
\subsection{Experimental Results}
Epitaxial RuO$_{2}$ films with similar thickness ($\sim10$ nm) were grown on TiO$_{2}$(100) and TiO$_{2}$(110) single crystal substrates using an oxide MBE. The details of the growth procedure were described elsewhere \cite{ruf2021strain, wadehra2025strain}. Figure S1 shows the reflection high-energy electron diffraction (RHEED) patterns recorded immediately after the growth of RuO$_{2}$. The RHEED results exhibit streaky patterns for the RuO$_{2}$ films grown on TiO$_{2}$(100) and TiO$_{2}$(110), suggesting smooth films. Figure S2 exhibits the X-ray diffraction results of RuO$_{2}$ films grown on the TiO$_{2}$(100) and TiO$_{2}$(110) substrates. The film thickness is calculated using the Laue fringes of $\theta$-2$\theta$ XRD scan. The $\theta$-2$\theta$ XRD scan only shows a single set orientation, confirming that the films are single-phase and oriented along the out-of-plane direction of the substrate. Figure S2 also shows the X-ray reciprocal space mapping (RSM), measured along two perpendicular in-plane directions of the substrate. RSM exhibits an anisotropic strain relaxation with films remaining coherent within the experimental error along one direction and partially relaxed along the highly strained direction, consistent with previous reports for similar thickness \cite{ruf2021strain, wadehra2025strain, jeong2025anisotropic}. Here, the lattice mismatch between RuO$_{2}$ film grown on TiO$_{2}$(100) is $-$4.7$\%$ along [001] and +2.2$\%$ along [010] (Figure 1b). The lattice mismatch between RuO$_{2}$ film grown on TiO$_{2}$(110) is $-$4.7$\%$ along [001] and +2.3$\%$ along [1$\bar1$0] (Figure 1c). The anisotropic strain gives rise to an anisotropic strain relaxation. RuO$_{2}$ films grown on TiO$_{2}$(110), for example, could remain coherent along [1$\bar1$0] up to 17 nm film thickness, while the relatively large compressive strain along [001] initiates relaxation with 4 nm critical thickness. 

Hall bars were fabricated using conventional lithography along the various crystallographic directions. Figure 2 shows the sheet resistance with current and temperature. The resistance is measured differentially at each temperature with sweeping current. Films grown on TiO$_{2}$(100) and TiO$_{2}$(110) exhibit emergent superconductivity, consistent with previous reports \cite{ruf2021strain, wadehra2025strain, uchida2020superconductivity}. We estimate the critical temperature ($T_c$) to be $\sim0.28$ K and $\sim0.55$ K for the films grown on TiO$_{2}$(110) and TiO$_{2}$(100) substrates, respectively, yielding Bardeen–Cooper–Schrieffer (BCS) superconducting gaps ($\Delta \approx 1.76 k_B T_c$) of 42~$\mu e$V and 83~$\mu e$V. Consistent with the critical temperature, the measured critical current is higher for the film grown on a TiO$_{2}$(100) substrate. 

The measured critical current is asymmetric which is reminiscent of the superconducting diode effect (SDE) \cite{PhysRevLett.99.067004, ando2020observation, nadeem2023superconducting}. Here, however, the reversed current sweep direction (Figure 2b and d) exhibits a completely reversed asymmetry in the critical current. Asymmetric critical current has been attributed to intrinsic (e.g., unconventional pairing) and extrinsic (e.g., asymmetric vortex dynamics and domain wall Josephson junctions) phenomena \cite{ma2025superconducting, PhysRevB.80.174517}.  

\begin{figure}[]
    \centering
    \includegraphics[width=1.\columnwidth]{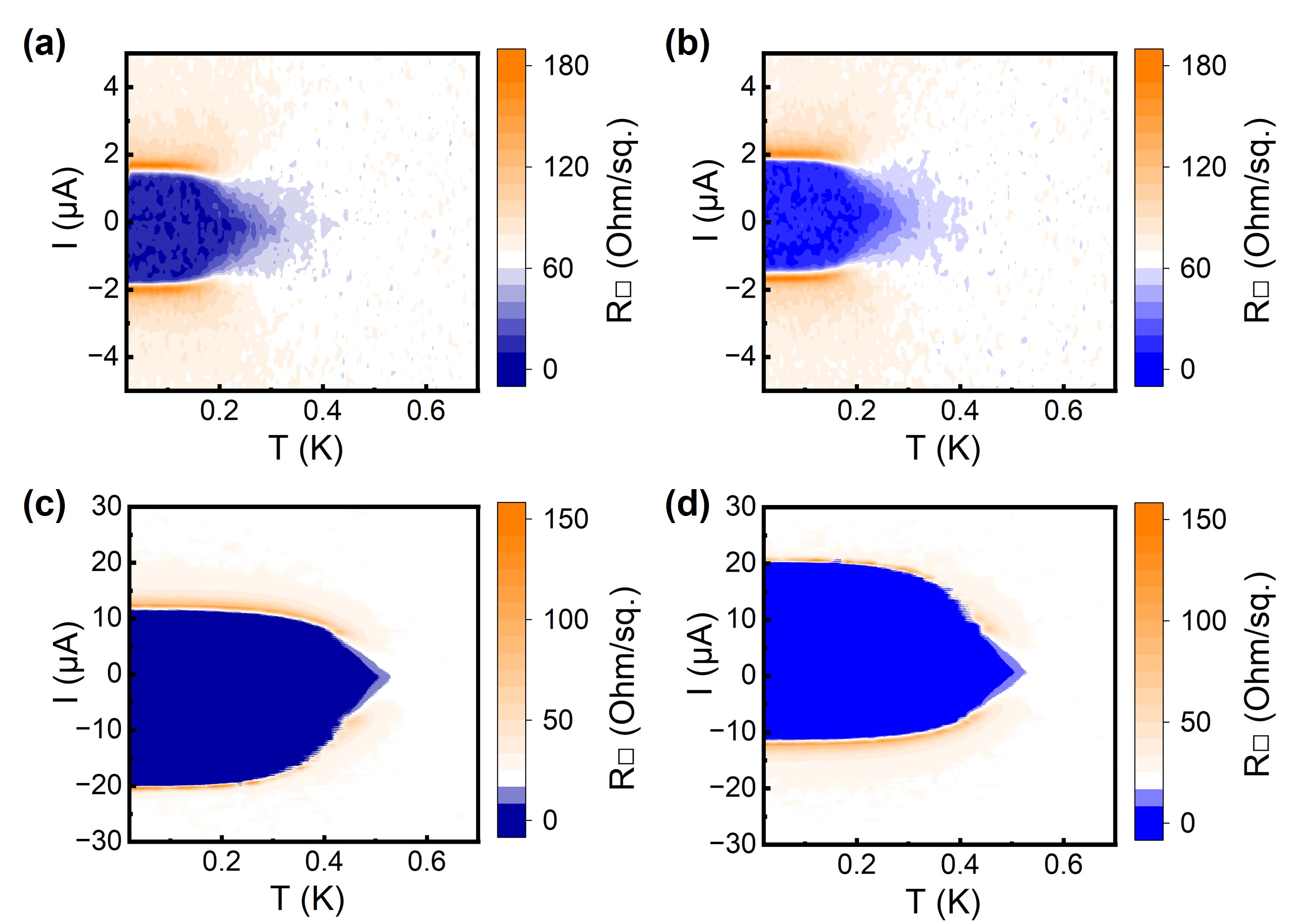}
 \caption{\textbf{Asymmetric superconducting transition of epitaxially strained RuO$_{2}$ films.} Sheet resistance with temperature and sweeping current for RuO$_{2}$ films grown epitaxially on (a) and (b) TiO$_{2}$(110), with current along [1$\bar1$1] and on (c) and (d) TiO$_{2}$(100), with current along [011]. The current sweeping direction is downward and upward in the left and right panels, respectively. The resistance is measured differentially at each temperature with sweeping current. Samples exhibit an abrupt superconducting transition with critical temperature ($T_c$) of $\sim0.28$ K and $\sim0.55$ K for films grown on TiO$_{2}$(110) and TiO$_{2}$(100) substrates. The film grown on TiO$_{2}$(100) substrate exhibits enhanced critical temperature and critical current compared to the film grown on TiO$_{2}$(110) substrate. The critical current is asymmetric. The asymmetry in the critical current reverses with sweeping direction of the current.}
    \label{fig:Fig2}
\end{figure}

The carrier mean free path is resolved to be $\approx 19$ nm and $\approx 36$ nm, $l_\mathrm{mfp}=\frac{h}{e^2}\frac{1}{K_f R_s}$, where $K_f=\sqrt{2\pi n_s}$, assuming a bulk carrier density ($n=8.87 \times 10^{21}$ cm$^{-3}$, \cite{peng2025universal}). We estimate the relaxation times as $\tau_\mathrm{tr}=m^* \mu/e$ with an effective mass of $m^*=2.4$, \cite{wu2025fermi}. We obtain $\tau_\mathrm{tr}=1.7 \times 10^{-14}$~sec and $\tau_\mathrm{tr}=3 \times 10^{-14}$~sec for films grown on TiO$_{2}$(110) and TiO$_{2}$(100) substrates, respectively. The calculated BCS superconducting gap ($\Delta \approx 1.76 k_B T_c$) yields $\Delta \tau_\mathrm{tr}/\hbar=0.001$ and $=0.004$ well in the dirty regime, pointing to spin-orbit scattering effects \cite{WHH1966, KLB1975}. 

\begin{figure*}[]
    \centering
    \includegraphics[width=2\columnwidth]{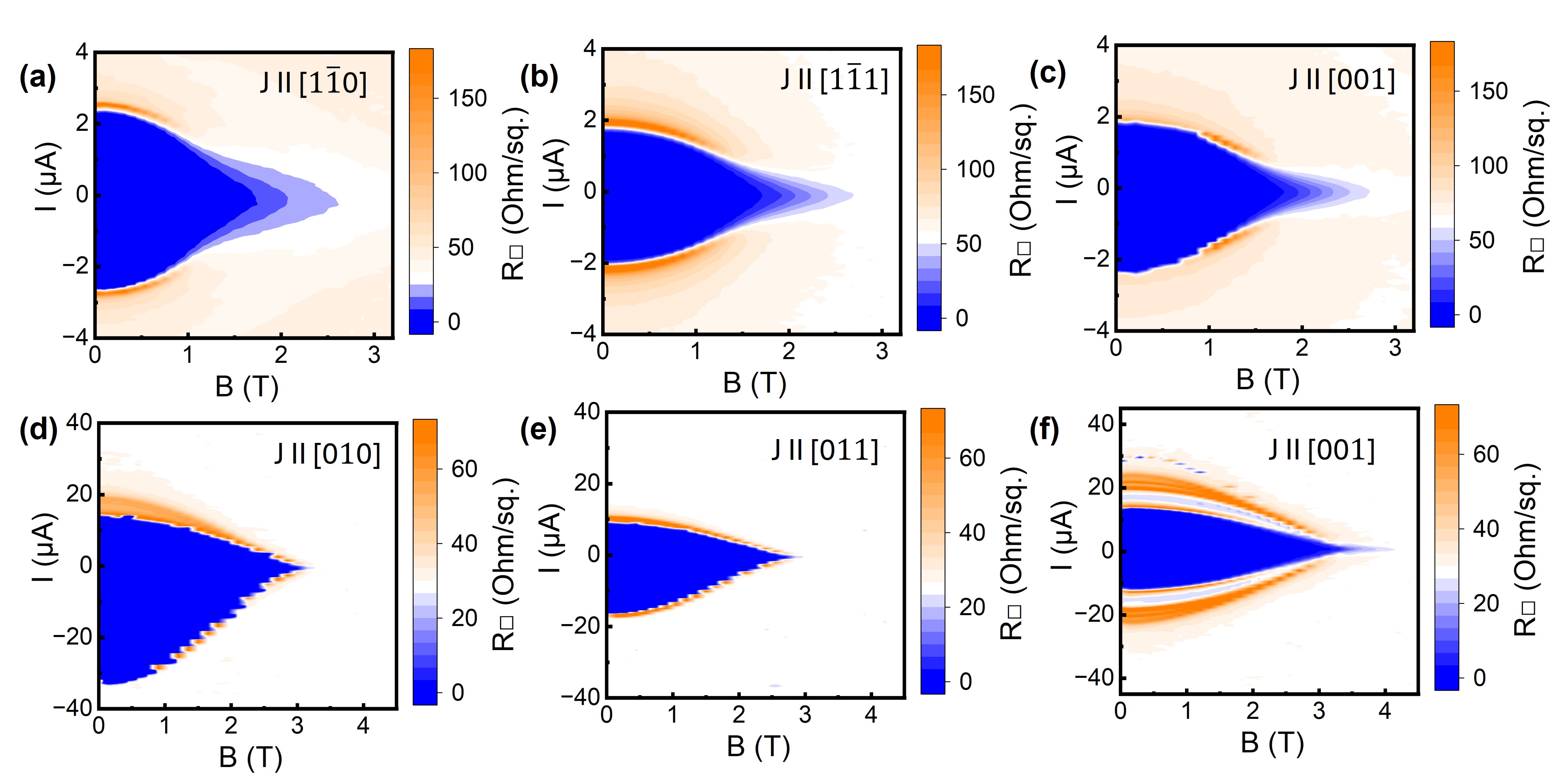}
 \caption{\textbf{Anisotropic superconducting transition with current and in-plane magnetic field.} Sheet resistance with applied in-plane magnetic field and sweeping current for RuO$_{2}$ films grown on TiO$_{2}$(110) and measured along (a) [1$\bar1$0], (b) [1$\bar1$1], and (c) [001]. Sheet resistance with applied in-plane magnetic field and sweeping current for RuO$_{2}$ films grown on TiO$_{2}$(100) and measured along (d) [010], (e) [011], and (f) [001]. Figure S3 shows a representative sample with fabricated Hall bars.}
    \label{fig:Fig3}
\end{figure*}

We further study the superconducting transition by measuring the longitudinal transport with applied in-plane and out-of-plane magnetic fields. Figure S3 exhibits the sheet resistance with current and applied out-of-plane magnetic field. Figure 3 shows the sheet resistance with current and applied in-plane magnetic field measured along various crystallographic directions. Here, the first observation is that the in-plane critical field ($H_{c2,\parallel}$) is larger for the film grown on TiO$_{2}$(100) substrate. This is consistent with the enhanced critical temperature and critical current in this film (Figure 2).

Next, we note the anisotropic nature of superconductivity where the magnitude of $H_{c2,\parallel}$ and asymmetry in the critical current depend on the crystallographic direction of the electronic transport. This anisotropy is more accentuated in the film grown on TiO$_{2}$(100) substrate. As discussed earlier the asymmetry observed in the critical current could be due to extrinsic effects such as defects that locally suppress the superconducting order parameter and pin vortices. Films grown on both TiO$_{2}$(100) and TiO$_{2}$(110) substrates experience a relatively large misfit which is anisotropic, as schematically illustrated in Figure 1. The anisotropic strain relaxation and its corresponding defects could give rise to an anisotropic superconducting transition and potentially explain the asymmetry in the critical current.

To further understand the nature of enhanced critical field we measured the sheet resistance with sweeping in-plane magnetic field at various temperatures (Fig. 4a and Fig. 4b). Using Figure 4 and Figure S3, we estimate the out-of-plane critical field ($H_{c2,\bot}$) and in-plane critical field ($H_{c2,\parallel}$) to be $H_{c2,\bot}=$0.3 ~T and $H_{c2,\parallel}=$2.8 ~T for the film grown on TiO$_{2}$(110) and $H_{c2,\bot}=$0.2 ~T and $H_{c2,\parallel}=$4 ~T for the film grown on TiO$_{2}$(100). Here, the $H_{c2,\parallel}/H_{c2,\bot}$ ratio is 9 and 20 for films grown on TiO$_{2}$(110) and TiO$_{2}$(100) substrates, respectively. 

To understand the nature of the Cooper pairs and evaluate their resilience against an in-plane magnetic field, the underlying pair-breaking mechanisms must be considered. In a superconductor, the upper critical field is constrained by two primary limiting factors: paramagnetic limiting and orbital pair breaking. The orbital mechanism originates from the Lorentz force, where the kinetic energy imparted to the electrons under an applied magnetic field exceeds the condensation energy, destroying the Cooper pairs. The orbital limit can also be framed as the penetration of magnetic flux quanta through the effective cross-sectional area of a Cooper pair ($\sim\xi^2$, where $\xi$ denotes the superconducting coherence length). For an in-plane magnetic field applied to a thin layer where the superconducting film thickness ($d_s$) is less than the coherence length ($d_s < \xi$), this effective cross-sectional area scales as $\sim d_s\xi$. Consequently, when the magnetic field is aligned parallel to the plane of a thin superconducting layer, the orbital pair-breaking effect is suppressed due to geometric restriction of the flux-threading area and the reduction of available out-of-plane momentum states.

Pauli paramagnetic pair breaking occurs due to the shift of the spin-down and spin-up electron energies. Assuming BCS coupling, the Pauli paramagnetic limit is estimated to be $H_P=\Delta/\sqrt{2}\mu_B$. Here, the Pauli limit, $H_P$, is estimated to be $\approx0.5$ ~T and $\approx1$ ~T for films grown on TiO$_{2}$(110) and TiO$_{2}$(100) substrates, respectively. We note the violation of Pauli limit in $H_{c2,\parallel}$. Figures 4a and 4b show the sheet resistance with sweeping in-plane magnetic field at various temperatures. Figures 4c and 4d exhibit the $H_{c2,\parallel}/H_p$ with temperature. We report base temperature $H_{c2,\parallel}/H_p$=5.5 and 4, for films grown on TiO$_{2}$(110) and TiO$_{2}$(100) substrates, respectively. The critical field monotonically increases as the sample is cooled. Enhanced in-plane critical field and violation of Pauli limit ($H_{c2,\parallel}/H_p$ as high as 9) has been reported recently in thin superconductors with enhanced spin-orbit coupling \cite{arnault2023, al2023enhanced, PhysRevB.111.214506}.

The out-of-plane critical field with temperature is described by the Ginzburg-Landau (GL) theory, $H_{c2,\bot}=\Phi_0(1-T/T_{c})/[2\pi(\xi_{GL})^2]$, where $\Phi_0$ and $\xi_{GL}$ are magnetic flux quantum and GL coherence length, respectively. We estimate the $H_{c2,\bot}(0)$ to be similar to the base temperature measurement of $H_{c2,\bot}$ (0.3 T and 0.2 T for films grown on TiO$_{2}$(110) and TiO$_{2}$(100), respectively). The GL fit using $H_{c2,\bot}(0)$ yields $\xi_{GL}$$\approx 33$ nm and $40$ nm for RuO$_{2}$ films grown on TiO$_{2}$(110) and TiO$_{2}$(100), respectively. The coherence length in both cases is larger than the film thickness ($\approx 10$ nm), suggesting a two-dimensional superconducting regime ($d_s<<\xi$).

\subsection{Mechanism for anisotropic superconductivity and enhanced critical field} 

First, we address the anisotropic superconducting transition. Here, the dependence of asymmetric critical current on the transport direction could point to underlying mechanism for anisotropic superconducting transition. Unconventional pairing, including finite momentum Cooper pairs, could give rise to asymmetric critical current \cite{PhysRevLett.128.037001, yuan2022supercurrent}. These unconventional pairing, however, typically require broken inversion and/or time reversal symmetries. Here, RuO$_{2}$ has rutile centrosymmetric crystal structure (P4$_{2}$/mnm). Furthermore, recent reports suggest a non-magnetic ground state for stoichiometric bulk and relaxed films \cite{PhysRevLett.132.166702, PhysRevLett.133.176401, kessler2024absence, gregory2025resonant}. We conclude that the observed critical current asymmetry is likely due to extrinsic effects. Here, the anisotropic epitaxial strain relaxation and its corresponding defects could be the main source of spatial non-uniformity that gives rise to these extrinsic effects.

The moving Abrikosov vortices in type II superconductors could be pinned by relaxation defects. The dissipation associated with vortex dynamics is asymmetric with respect to the current direction. Rectification effects, observed in thin superconducting layers with nano-scale defects, have been attributed to vortex dynamics \cite{PhysRevB.76.220507, de2006controlled, PhysRevLett.95.087002, PhysRevB.85.012502, PhysRevB.81.174505, PhysRevLett.106.137003, PhysRevB.73.014507, PhysRevLett.94.057003}. Here, however, the asymmetric critical current is present even in the absence of applied magnetic field, suggesting that vortex dynamics does not play a major role in the observed critical current asymmetry. The relaxation defects could also suppress the superconducting order parameter locally, giving rise to an array of Josephson junctions. Josephson junctions are known to be a source of asymmetric critical current and SDE \cite{PhysRevLett.99.067004, ando2020observation, nadeem2023superconducting, carapella2009bistable, golod2022demonstration, rashidi2024vortex, PhysRevLett.94.057003}. Superconductivity in epitaxially strained RuO$_{2}$ films does not emerge in thinner layers, where the film remains coherent along various directions, and in thicker layers ($>$50 nm) due to relaxation of the epitaxial strain which is necessary for the emergence of superconductivity \cite{ruf2021strain, wadehra2025strain}.

To understand the nature of anisotropic strain relaxation and its corresponding defects, We carried out hard x-ray diffraction experiments in the film grown on TiO$_{2}$(100) which shows enhanced anisotropy. Figure S4 exhibits two perpendicular slices through the reconstructed volume around the 110 Bragg peak: a K–L slice at constant H (Figure S4a) and an H–L slice at constant K (Figure S4b). Faint streaks run along 101 and $\bar1$01 in reciprocal space, indicating planar defects parallel to the (101) and ($\bar1$01) planes in real space. The streaks are narrow along K, implying that these defects extend along the b axis. No streaks are visible in the K–L slice, suggesting absence of similar defects in the (0$\bar1$1) and (011) planes. This is consistent with the transmission electron microscopy observations, reporting planar defects along the (101) and ($\bar1$01) planes in similarly grown films\cite{wadehra2025strain}. Here, the diffraction results confirm the presence of anisotropic relaxation planar defects which could explain the observed anisotropy of superconducting transition.

\begin{figure}[]
    \centering
    \includegraphics[width=1\columnwidth]{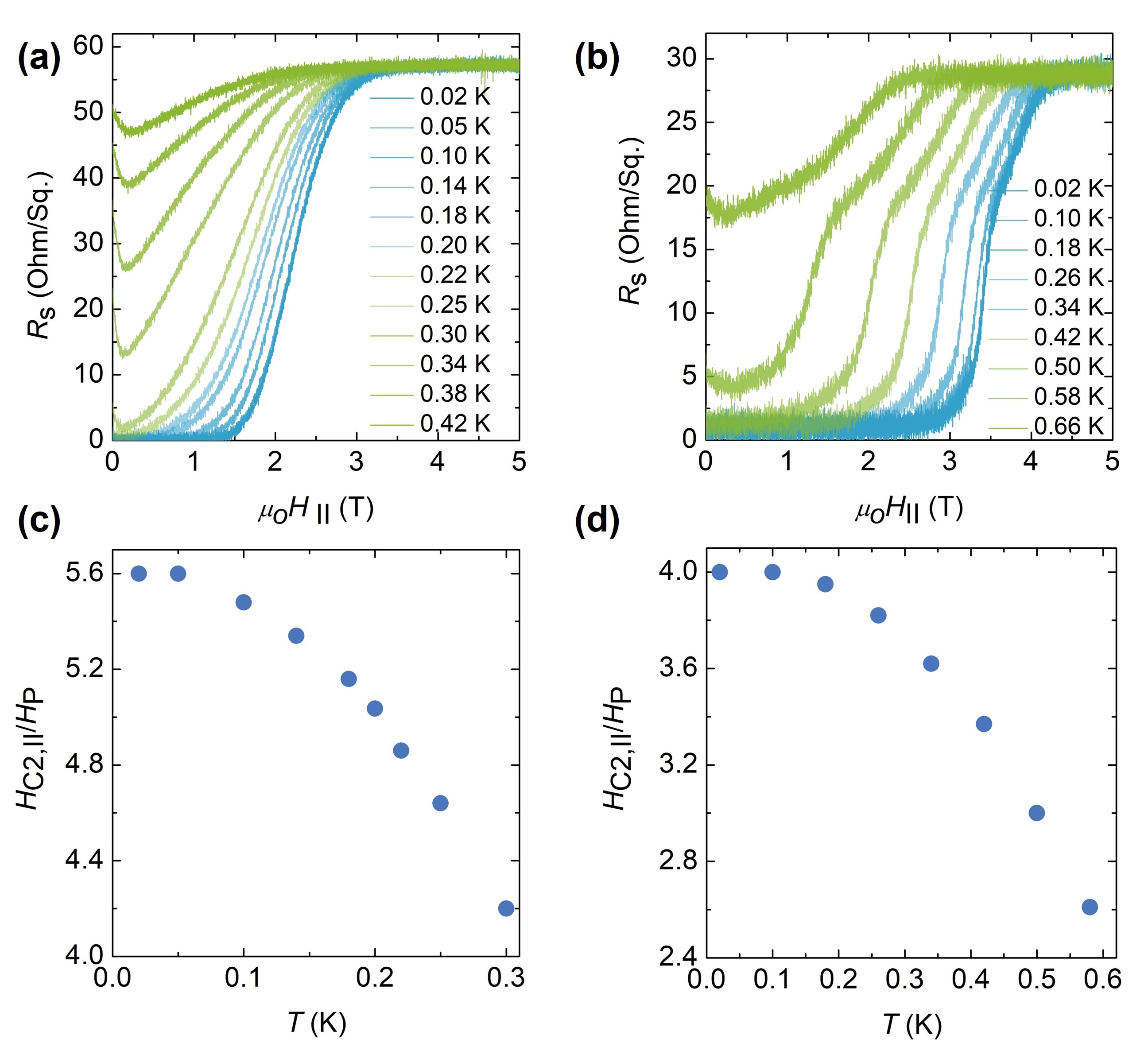}
 \caption {\textbf{Violation of Pauli paramagnetic limit in epitaxially strained RuO$_{2}$ films.} Sheet resistance with sweeping in-plane magnetic field at various temperatures for RuO$_{2}$ films grown on (a) TiO$_{2}$(110)  and (b) TiO$_{2}$(100) single crystal substrates, respectively. The measured in-plane critical field violates Pauli limit with $H_{c2,\parallel}/H_p$=5.5 and 4 at base temperature for RuO$_{2}$ films grown on (c) TiO$_{2}$(110)  and (d) TiO$_{2}$(100) single crystal substrates, respectively. Here, the measured $H_{c2,\parallel}$ is defined at 0.9$R_n$, where $R_n$ is the normal state resistance. The Pauli paramagnetic limit is estimated using $H_P=\Delta_0/\sqrt{2}\mu_B$ relationship, where $\Delta \approx 1.76 k_B T_c$.}
    \label{fig:Fig4}
\end{figure}

Next, we discuss the enhanced $H_{c2,\parallel}$ and violation of Pauli limit. Two various mechanisms determine $H_{c2,\parallel}(0)$. In the case of orbital currents determining the critical field, a GL estimate yields
$H_{c2,\parallel}(0)=\Phi_0 \sqrt{3}/(\pi d_{sc}\xi_{GL})$, where $d_{sc}$ is the effective thickness of the superconducting layer. We estimate the $H_{c2,\parallel}(0)$ to be similar to the base temperature measurement of $H_{c2,\parallel}$ (2.8 T and 4 T for films grown on TiO$_{2}$(110) and TiO$_{2}$(100), respectively). Using the above values we obtain $d_{sc}\sim12$ nm and $\sim7$ nm for films grown on TiO$_{2}$(110) and TiO$_{2}$(100) substrates, respectively. The calculated $d_{sc}$ does not match the $\approx10$ nm film thickness and results in unphysical $d_{sc}$ for the film grown on TiO$_{2}$(100).

Superconductivity can also be quenched at a first-order transition at the Pauli paramagnetic limit. Under the assumption of weak coupling, we can derive the critical field by equating the free energies of the normal and superconducting states given by, 

\begin{equation}
    H_{c2,\parallel}= H_P \sqrt{\frac{\chi_P}{\chi_N-\chi_{SC}}},\label{eq:pauli} 
\end{equation}
where $\chi_P=\mu_B^22N_F$ is the Pauli spin susceptibility of non-interacting electrons with density of states per spin at the Fermi level $N_F$, and $\chi_{N(SC)}$ is the normal-state (superconducting) magnetic susceptibility. For singlet superconductivity and in the absence of spin-orbit coupling $\chi_{SC}=0$, $\chi_N=\chi_P$ and $H_{c2,\parallel}(0)=H_p=  \frac{\Delta}{\sqrt{2}\mu_B} \approx0.5$~T for the film grown on TiO$_{2}$(110) and $\approx1$~T for the film grown on TiO$_{2}$(100).

Here, however, we report in-plane critical field considerably exceeding the Pauli limit ($H_{c2,\parallel}/H_p$=5.5 and 4, for films grown on grown on TiO$_{2}$(110) and TiO$_{2}$(100) substrates, respectively). Unconventional pairing could result in enhanced critical field beyond Pauli limit. For example, Pauli limit does not apply to $p$-wave superconductors and enhancement of in-plane critical field with mixed parity has been reported with broken inversion symmetry \cite{schumann2020possible}. Here, the inversion and time-reversal symmetries are expected to remain intact \cite{PhysRevLett.132.166702, PhysRevLett.133.176401, kessler2024absence, gregory2025resonant}. 

Another possibility is inversion symmetry breaking at the interface/surface of the film, giving rise to an admixture of triplet superconductivity. This will contribute to a maximum critical field enhancement with respect to $H_P$ of  $\sqrt2$ with Rashba spin-orbit coupling \cite{gor2001superconducting}, which does not explain the measured $H_{c2,\parallel}/H_p$ $\sim 5.5 $. 

Fulde–Ferrell–Larkin–Ovchinnikov (FFLO) state could also enhance the critical field \cite{fulde1964superconductivity}. The FFLO state emerges in the clean regime $\Delta \tau_\mathrm{tr}/\hbar<<1$ whereas here, as discussed above,  the system is deep in the dirty regime with mean free path ($l_\mathrm{mfp}$$\approx 19$ nm and $36$ nm) smaller than the superconducting coherence length ($\xi_{GL}$$\approx 33$ nm and $40$ nm). 

In the dirty limit, strong spin-orbit scattering due to quenched disorder could enhance the upper critical field. This scattering randomizes electron spins, effectively screening the polarizing effect of the applied magnetic field \cite{WHH1966,Hwang2012,Hwang2018,lu2014}. Furthermore, it is reported that the heavy element incorporation in thin films enhances $H_{c2,\parallel}$, due to weakening of the polarizing effect of the magnetic field with spin-orbit scattering \cite{KLB1975}. Here, the itinerant charge carriers are from ruthenium 4$d$ $t_{2g}$ manifold which has sizable spin-orbit coupling. 

\section*{Conclusions}

In summary, our experimental results demonstrate an anisotropic two-dimensional superconductivity emerging in thin layers of RuO$_{2}$ grown epitaxially on TiO$_{2}$(110) and TiO$_{2}$(100) single crystal substrates. The measured in-plane critical fields ($H_{c2, \parallel}$) are well beyond the Pauli limit. Here, the anisotropic nature of superconductivity is attributed to anisotropic epitaxial strain and its associated anisotropic relaxation defects. Our results also show that spin-orbit scattering is essential to describe the enhancement of the critical field. These results offer opportunities for epitaxial control and engineering of superconductors.

\section*{Data Availability Statement}
 
The data that support the findings of this study are available in the article and its Supplemental Material. Raw data can be obtained from the corresponding authors upon request.

\begin{acknowledgments}
K.A acknowledges discussions with Mohit Randeria and Peter Anderson. Y.G and S.J.P were supported by the U.S. National Science Foundation under Grant No. NSF DMR-2408890. B.Z.G. and A.S. acknowledge support by the Department of Energy – Office of Basic Energy Sciences under award DE-SC0019414 (synchrotron x-ray characterization). Research conducted at the Center for High-Energy X-ray Sciences (CHEXS) is supported by the National Science Foundation (BIO, ENG and MPS Directorates) under award DMR-2342336.

\end{acknowledgments}

\section*{Appendix: Materials and methods}

Strained RuO$_{2}$ thin films were grown using a Veeco GEN10 molecular-beam epitaxy (MBE) system on TiO$_{2}$ (110) and (100) substrates. Prior to growth, TiO$_{2}$ substrates were cleaned with organic solvents, etched in acid, and annealed in air to produce starting surfaces with step-terrace morphology \cite{yamamoto2005preparation}. A molecular beam of ruthenium, 99.99\% pure ruthenium metal (ESPI Metals), was generated using an electron beam evaporator. A background pressure of 1×10$^{-6}$ torr of distilled ozone (~80$\%$ O$_{3}$ + 20$\%$ O$_{2}$) was used as the oxidant, and a thermocouple in close proximity to the sample was used to measure the temperature. The growth details were described elsewhere \cite{ruf2021strain, wadehra2025strain}. In-situ RHEED patterns were recorded using KSA-400 software and a Staib electron source operated at 13 kV and a filament current of 1.55 A. X-ray diffraction, x-ray reflectometry, and reciprocal space mapping measurements were carried out using a PANalytical Empyrean diffractometer with Cu K$\alpha$1 radiation.

The sub-Kelvin magnetoelectric measurements were carried out in an Oxford dilution refrigerator with base temperature of 10 mK as measured by Cernox sensors (50 mK with the application of high field). This refrigerator is equipped with a 16T superconducting magnet, which provides the magnetic field in the experiments. All electrical lines are fitted with QDevil filters, which comprises of 3-stage low frequency RC filters. Longitudinal resistance R$_{xx}$ are measured using Hall bar geometry by sourcing at 7.6 Hz.

The large-volume reciprocal-space maps were collected at the QM2 (Q-mapping for Quantum Materials) Beamline at ID4B, CHESS. While the sample was rotated 360° in azimuth across 3600 steps, a Pilatus 6M area detector recorded a slice of the Ewald sphere at each frame. An in-house code was used to interpolate the measured intensities onto a uniform reciprocal-space grid. 

First-principles electronic structure calculations based on density functional theory (DFT) 
were performed using the Quantum ESPRESSO software package \cite{giannozzi2009quantum, giannozzi2017advanced}. The generalized gradient approximation as implemented in the PBEsol functional \cite{perdew2008restoring} was employed as the exchange-correlation functional with norm-conserving pseudopotentials and plane-wave basis sets, using a kinetic energy cutoff of Ry, an electronic momentum k-point mesh of $16 \times 16 \times 24$, 
Methfessel-Paxton smearing of 20 meV for the occupation of the electronic states, and a tolerance of $10^{-10}$ eV for the total energy convergence. The bulk lattice parameters of RuO$_2$ calculated with the PBEsol functional are $a = 4.464$ \AA, and $c = 3.093$ \AA. The strained RuO$_2(100)$ structure is calculated by changing the lattice constants of this bulk crystal by $+2.2\%$ along $[010]$ and $-4.7\%$ along the $[001]$-axis while relaxing the structure along the $[100]$-axis. The strained RuO$_2(110)$ structure is calculated by changing the lattice constants of this bulk crystal by $+2.2\%$ along $[\bar110]$ and $-4.7\%$ along the $[001]$-axis while relaxing the structure along the $[110]$-axis. The projected density of states were calculated with the Wannier90 code \cite{pizzi2020wannier90}, and the coordinate system for the orbital projections were reoriented to be aligned with Ru-O bonds.

\bibliographystyle{apsrev4-2}
\bibliography{apssamp}


\end{document}